\documentclass[onecolumn,superscriptaddress, aps,pre]{revtex4-2}
\makeatletter
\newcommand*{\rom}[1]{\expandafter\@slowromancap\romannumeral #1@}
\makeatother
\usepackage{color}
\usepackage{graphicx}
\usepackage{hyperref}
\usepackage{amsmath,amssymb}
\usepackage{epstopdf}
\usepackage{subcaption}
\graphicspath{{figs/}}
\def\beq{\begin{equation}}
\def\eeq{\end{equation}}
\def\bea{\begin{eqnarray}}
\def\eea{\end{eqnarray}}

\begin{document}
\title{$q$-Deformed Gross Pitaevskii Equation}
\author {Mahnaz Maleki}
\email{m.maleki@uma.ac.ir}
\affiliation{Department of Physics, University of Mohaghegh Ardabili, P.O. Box 179, Ardabil, Iran}
\author {Hosein Mohammadzadeh}
\email{mohammadzadeh@uma.ac.ir}
\affiliation{Department of Physics, University of Mohaghegh Ardabili, P.O. Box 179, Ardabil, Iran}
\begin{abstract}
We derive the Gross Pitaevskii equation (GPE) for condensate of bosons obeying deformed statistics under external potential and inter-particle interaction. First, we obtain the well-known Schrodinger equation. Using a suitable Hamiltonian for condensate phase and minimizing the free energy of the system, we find out the $q$- deformed GPE. Thus, at very low temperature, where the dynamics of excited-occupation level can be neglected, the dynamics of a deformed statistics system can be described by the GPE, similar to the Bose-Einstein condensate.
\end{abstract}
\maketitle
%%%%%%%%%%%%%%%%%%%%%%%%%%%%%%%%%%%%%%%%%%%%%%%%%%%
\section{Introduction}\label{1}
Athough the known particle statistics, which is include Maxwell Boltzmann, Fermi Dirac and Bose Einstein, some generalized statistics also have been proposed based on experimental observations,  including anyons  which are particles whose wave function under the exchange of two particles  takes a non-trivial phase\cite{wilczek1991anyons,wilczek1992disassembling,mirza2010thermodynamic}.  Different deformed statistics correspond to the different deformations of commutative and anti-commutative algebras such as $q$-fermions (and bosons)\cite{lavagno2002generalized,tuszynski1993statistical,mirza2011thermodynamic}, $qp$-deformed statistics\cite{daoud1995statistical,mohammadzadeh2017thermodynamic}. Also, the fractional exclusion and exchange statistics have been used to describe the fractional quantum Hall effect \cite{stern2008anyons,lee1989anyon,tevosyan1997virial}. Other kinds of fractional exclusion statistics are Gentile statistics\cite{gentile1940itosservazioni} and Polychronakos\cite{polychronakos1996probabilities}.

There are many suggestions which indicate that the generalized statistics may indeed be the statistics of certain cosmological objects. The properties of dark matter and dark energy have been described using the $q$-deformed statistics \cite{maleki2022description,maleki2022mass}, infinite statistics\cite{ebadi2013infinite}, $\mu$- deformed\cite{gavrilik2018condensate}, non-extensive statistics\cite{barboza2015dark,beck2004nonextensive,leubner2005nonextensive,zadeh2018note}, ewkons statistics\cite{hoyuelos2018exotic,hoyuelos2018creation}. Also the maximum available temperature of a system with particles that obey q-deformed statistics have been investigated. 

The correct dynamics of a real particle system obeying  Bose-Einstein statistics, include the external potential and interparticle interaction. For such system, the dynamics can be described by the GPE\cite{rogel2013gross}. The GPE is a nonlinear equation and has many utilization in different branches of physics specially in Bose-Einstein condensation (BEC) statistics. We can mention the successful experimental finding of BEC of dilute trapped bosonic alkali-metal atoms 7Li, 23Na, and 87Rb at ultra-low temperatures\cite{pitaevskii2016bose,leggett2001bose,pethick2008bose}, the BEC in optical and quasi-crystalline optical lattices \cite{niu2020bose,hu2020vortices,wang2021dynamics}, the spinor BEC \cite{yin2017solitons,hu2020vortices}.
Furthermore, the time dependent GPE explains the dynamics of initially trapped BEC \cite{erdHos2007rigorous,benedikter2015quantitative,pickl2015derivation,brennecke2019gross}.
There are various solutions for the GPE such as analytical \cite{liu2017analytical,trallero2008formal,yu2018localized}, numerical \cite{antoine2018numerical,sataric2016hybrid,vergez2016finite,vudragovic2012c}, soliton solutions\cite{su2016nonautonomous,gravejat2015asymptotic,bethuel2008orbital,liu2017dark,zakeri2018solitons,yin2017solitons,yu2019novel} and stationary \cite{charalampidis2018computing,charalampidis2020bifurcation}. Also, some other kinds have been developed of solutions the equation, such as the inverse scattering method\cite{yu2019inverse}. Solutions of the GPE with various potentials have been investigated for various potentials such as the external potentials  \cite{yu2018localized,li2017non}, the changed external trap\cite{pickl2015derivation}, the nonlinear lattice pseudo-potential \cite{alfimov2019localized}, a sort of parity-time-symmetric potentials \cite{barashenkov2016exactly}, the multi-well potential \cite{guo2018properties} and a parabolic potential \cite{liu2020new}. Of course, the GPE is used to study  in some other aspects of dark matter such as the velocity of rotation, density and  mass profile of galaxies \cite{harko2011cosmological,harko2011bose,harko2015testing,zhang2018slowly}. Furthermore, BEC has been utilized in some characteristics of cosmology  and gravitation including  the gravastar (a de Sitter star) of black hole physics \cite{cunillera2018gross} and the black hole in the anti de Sitter space \cite{biasi2017delayed}.

So far in paper, we have introduced the applications of the GPE for BEC. For statistics deformed version of GPE is not yet known. In current paper, we derive the time-independent GPE of $q$-deformed statistics which can also be useful for research on various topics in physics. We show that if one only considers the ground state, the dynamics of a system with particle obeying q-deformed statistics can be described by the $q$-deformed GPE.

We organize the paper as follows: In section \ref{2}, we review the $q$-deformed statistics. In section\ref{3}, we obtain the time-independent  Schrodinger equation in three dimension for $q$- deformed formalism. Also, in section\ref{4}, we calculate the time-independent  $q$- deformed GPE. We conclude the paper in section \ref{5}.

%%%%%%%%%%%%%%%%%%%%%%%%%%%%%%%%%%%
%%%%%%%%%%%%%%%%%%%%%%%%%%%%%%%%%%%%%%%%%
\section{q-OSCILLATOR ALGEBRA AND q-CALCULUS }\label{2}
In this section, we briefly review the basic aspects of $q$-oscillator algebra of creation and annihilation operators introduced by Biedenharn and McFarlane\cite{biedenharn1989quantum,macfarlane1989q} and the main properties of $q$- calculus and basic- deformed elementary functions which are useful in the present investigation. The symmetric $q$-oscillator algebra is defined in terms of the creation operator ($c^\dag$), annihilation operator ($c$) and the $q$-number operator ($N=c^\dag c$) as follows
 \bea
cc^{\dag}-kqc^{\dag}c=q^{-N},~~~~~~ [c,c]_{k}=[c^{\dag},c^{\dag}]_{k}=0,
\label{q-a}
 \eea
 \bea
[N,c^{\dag}]=c^{\dag},~~~~~~[N,c]=-c,
\label{q-b}
 \eea
 \bea
c^{\dag}c=[N],~~~~~~~~~~~cc^{\dag}=[1+k N],
\label{q-c}
 \eea
where $[x,y]_k=xy-kyx$ and $k=1 (k=-1)$ is related to $q$-bosons ($q$-fermions). The $q$-Fock space spanned by the orthornormalized eigenstates $|n\rangle$, corresponding to
\bea
|n\rangle=\frac{(c^{\dag})^{n}}{\surd[n]!}|0\rangle
\label{eigenstate}
\eea
where the $q$-basic factorial is defined as follows
\bea
[n]!=[n][n-1]...[1],
\label{factorial}
\eea
and the so-called $q\leftrightarrow q^{-1}$ symmetric basic number $[x]$ is defined in terms of the $q$-deformation parameter as follows
\bea
[x]=\frac{q^{x}-q^{-x}}{q-q^{-1}}.
\label{q-d}
\eea
In the limit of $q\rightarrow1$, the basic number $[x]$ reduces to the ordinary number $x$ and all the above relations reduce to the standard boson relations.

The actions of $c$, $c^{\dag}$ on the Fock state $|n\rangle$ are given by
\bea
c^{\dag}|n\rangle&=&[n+1]^{1/2}|n+1\rangle,\nonumber\\
c|n\rangle&=&[n]^{1/2}|n-1\rangle,\nonumber\\
N|n\rangle&=&n|n\rangle.
\label{q-aa}
\eea
The transformation from Fock observables to the configuration space (Bargmann holomorphic representation) can be accomplished by the replacement\cite{finkelstein1996q,lavagno2006classical,ubriaco1992non}
\bea
&&c^{\dag}\to x \ ,\\
&&c\to {\cal{D}}^{(q)}_{x} ,
\label{q-ab}
\eea
where ${\cal{D}}^{(q)}_{x}$ is the Jackson Derivative (JD)\cite{jackson1909generalization} which is defind as follows
\bea
{\cal{D}}^{(q)}_{x}=\frac{{\cal{D}}_{x}-({\cal{D}}_{x})^{-1}}{(q-q^{-1})x},
\label{q-ac}
\eea
where $D^{(q)}_x=q^{x\,\partial_x}$ is the dilatation operator. Its action on an arbitrary real function $f(x)$ is given as follows
\bea
{\cal{D}}^{(q)}_{x}f(x)=\frac{f(q x)-f(q^{-1}x)}{(q-q^{-1})x}.
\label{q-e}
\eea
Unlike the ordinary derivative, which measures the rate of change of the function in terms of an incremental translation of its argument, the JD measures its rate of change with respect a dilatation of its argument by a factor of $q$. The JD satisfies some simple proprieties that will be useful in the following. For instance, if JD acts on monomials $f(x)=x^n$ where $n\geq0$, we obtain that
\bea
\partial_{x}^{q} x^{n}=[n]_{q}x^{n-1}.
\label{q-f}
\eea
Moreover, we have the $q$-Leibnitz rule as follows\cite{lavagno2009quantum}
\bea
&{\cal{D}}^{(q)}_{x}\left(f(qx)g(x)\right)&=\left({\cal{D}}^{(q)}_{x}f(qx)\right)g(qx)+f(x){\cal{D}}^{(q)}_{x}g(x),\nonumber\\
&{\cal{D}}^{(q)}_{x}\left(f(q^{-1}x)g(x)\right)&=\left({\cal{D}}^{(q)}_{x}f(q^{-1}x)\right)g(q^{-1}x)+f(x){\cal{D}}^{(q)}_{x}g(x),
\label{q-g}
\eea
also, the follwing property holds
\bea
{\cal{D}}^{(q)}_{\beta x}g(x)=\frac{1}{\beta}{\cal{D}}^{(q)}_{x}g(x).
\label{q-h}
\eea
where $\beta$ is a constant. We will use of the properties of Eq. (\ref{q-g}) and Eq. (\ref{q-h}) in next sections.
%%%%%%%%%%%%%%%%%%%%%%%%%%%%%%%%%%%%%%%%%%%%%%%%%%%%%%%%%%%%%%%%%%%%%%%%%%%%%%%
%%%%%%%%%%%%%%%%%%%%%%%%%%%%%%%%%%%
\section{q-DEFORMED QUANTUM MECHANICS IN THREE DIMENSION}\label{3}
%%%%%%%%%%%%%%%%%%%%%%%%%%%%
The coordinate of  $q$-deformed position and  the coordinate of $q$-deformed momentum are defined as follows
\bea
\hat{x}=x,~~\hat{p}=\frac{\hbar}{i}{\cal{D}}_{x}^{2(q)}.
\label{x-p}
\eea
Then, using Eq.(\ref{x-p}) can defined the time-independent Schrodinger equation in three dimension as follows
\bea
\left[-\frac{\hbar^{2}}{2m}\left({\cal{D}}_{x}^{2(q)}+{\cal{D}}_{y}^{2(q)}+{\cal{D}}_{z}^{2(q)}\right)+V_{q}(x,y,z)\right]\psi(x,y,z) =E\psi(x,y,z) 
\label{q-deform}
\eea
%Now we can calculate the sentence ${\cal{D}}^{q^{2}}_{r}\psi(x,y,z)=\left({\cal{D}}^{(q)^{2}}_{x}+{\cal{D}}^{(q)^{2}}_{y}+{\cal{D}}^{(q)^{2}}_{z}\right)\psi(x,y,z)$. 
First, we calculate ${\cal{D}}_{x}^{2(q)}\psi(x)$ using the Eq.(\ref{q-e}) as follows
\bea
{\cal{D}}_{x}^{2(q)}\psi(x)={\cal{D}}^{(q)}_{x}\left(\frac{\psi(q x)-\psi(q^{-1}x)}{(q-q^{-1})x}\right)=(\frac{1}{q-q^{-1}})\left({\cal{D}}^{(q)}_{x}(\frac{\psi(q x)}{x})-{\cal{D}}^{(q)}_{x}(\frac{\psi(q^{-1}x)}{x})\right),
\label{q-l}
\eea
we consider $l(x)=\frac{1}{x}$, then we will  have the above eqution as follows
\bea
{\cal{D}}_{x}^{2(q)}\psi(x)=(\frac{1}{q-q^{-1}})\left({\cal{D}}^{(q)}_{x}(\psi(q x)l(x))-{\cal{D}}^{(q)}_{x}(\psi(q^{-1}x)l(x))\right),
\label{q-m}
\eea
using the $q$-deformed Liebnitz rule of Eq.(\ref{q-g}) and after simplification,the following equation is easily deduced as follows
\bea
{\cal{D}}_{x}^{2(q)}\psi(x)=\frac{1}{(q-q^{-1})}\left({\cal{D}}^{(q)}_{x}(\psi(q x))l(qx)-{\cal{D}}^{(q)}_{x}(\psi(q^{-1}x))l(q^{-1}x)\right),
\label{q-n}
\eea 
also using Eq. (\ref{q-h}), we obtain that
\bea
{\cal{D}}_{x}^{2(q)}\psi(x)=
 \frac{1}{(q-q^{-1})^{2}x} \left( \psi(q^{2}x)l(qx)-\psi(x)l(qx)-\psi(x)l(q^{-1}x)+\psi(q^{-2}x)l(q^{-1}x) \right).
\label{q-o}
\eea
We substitute $l(qx)=\frac{1}{qx}$ and $l(q^{-1}x)=\frac{1}{q^{-1}x}$ in the above equation and finally obtain that
\bea
{\cal{D}}_{x}^{2(q)}\psi(x)=\frac{q \left( \psi(q^{-2}x)-\psi(x) \right)-q^{-1} \left(\psi(q^{2}x)-\psi(x) \right)}{(q-q^{-1})^{2}x^{2}}.
\label{q-p}
\eea
As the some manner, we will derive the deformed three dimensional Laplacian operator as follows
\bea
({\cal{D}}_{x}^{2(q)}+{\cal{D}}_{y}^{2(q)}+{\cal{D}}_{z}^{2(q)})\psi(x,y,z)&=&
\frac{q}{(q-q^{-1})^{2}}\left(\frac{\psi(q^{-2}x)}{x^{2}}+\frac{\psi(q^{-2}y)}{y^{2}}+\frac{\psi(q^{-2}z)}{z^{2}}\right)\nonumber\\
&-&\frac{q^{-1}}{(q-q^{-1})^{2}}\left(\frac{\psi(q^{2}x)}{x^{2}}+\frac{\psi(q^{2}y)}{y^{2}}+\frac{\psi(q^{2}z)}{z^{2}}\right)\nonumber\\
&-&\frac{1}{(q-q^{-1})}\left(\frac{\psi(x)}{x^{2}}+\frac{\psi(y)}{y^{2}}+\frac{\psi(z)}{z^{2}}\right).
\label{Q-D}
\eea
Therefore, by substituting Eq.(\ref{Q-D}) into Eq.(\ref{q-deform}), we obtain the time-independent $q$-deformed schrodinger equation in three dimension as follows
\bea
&-&\frac{\hbar^{2}}{2m}\left[\frac{q}{(q-q^{-1})^{2}}\left(\frac{\psi(q^{-2}x)}{x^{2}}+\frac{\psi(q^{-2}y)}{y^{2}}+\frac{\psi(q^{-2}z)}{z^{2}}\right)
-\frac{q^{-1}}{(q-q^{-1})^{2}}\left(\frac{\psi(q^{2}x)}{x^{2}}+\frac{\psi(q^{2}y)}{y^{2}}+\frac{\psi(q^{2}z)}{z^{2}}\right) \right.\nonumber\\
&-&\left. \frac{1}{(q-q^{-1})}\left(\frac{\psi(x)}{x^{2}}+\frac{\psi(y)}{y^{2}}+\frac{\psi(z)}{z^{2}}\right)\right]+V_{q}(x,y,z)=E\psi(x,y,z)
\label{q-deform schrodinger equation}.
\eea
%%%%%%%%%%
\section{THE GROSS-PITAEVSKII EQUATION of DEFORMED STATISTICS}\label{4}
%%%%%%%%%%%%
For a condensed system in deformed thermodynamics, we consider all particles are in the same single particle state $\phi({\bf r})$ with $\int d^{3}{\bf r}|\phi({\bf r})|^{2}=1$. Thus, we have a system with $N$ particles with wavefunction $\Psi({\bf r}_{1},{\bf r}_{2}...,{\bf r}_{N})$ that is given by product of single-particle awavefunction $\phi$ as follows
\bea
\Psi({\bf r}_{1},{\bf r}_{2},...,{\bf r}_{N})=\prod_{i=1}^{N} \phi({\bf r}_{i}).
\label{function}
\eea
We start the description of the Hamiltonian in terms of the Kinetic energy, the external potential energy $(V_{ext})$ and the interaction potential between the $N$ Particles. We choose an interaction potential between bi-particles as follows
 \bea
V(|{\bf r_{i}}-{\bf r}_{j}|)=\frac{4\pi\hbar^2}{m} a\delta({\bf r_{i}}-{\bf r_{j}}),
\label{the interaction potential}
\eea
which $(N-1)$ particles interact by one of them and $a$ is the $s$-wave scattering length. 
%This wavefunction $\Psi({\bf r}_{1},{\bf r}_{2}...,{\bf r}_{N})$ is related to the Kinetic energy of the  particles, the external potential energy $(V_{ext})$ and the interaction between the $N$ Particles .
Thus, the  general Hamiltonian is given by
\bea
\hat H=\sum_{i=1}^N\left(\frac{{\bf p}_i^2}{2m}+V_{ext}({\bf r}_i)\right) + \frac{1}{2}\sum_{i=1}^{N}\sum_{j\neq i}^N V\left(|{\bf r}_i-{\bf r}_j|\right).
\label{HAMILTONIAN}
\eea
Evaluation of the wavefunction of the $q$-defrmed condensed phase using the above general Hamiltonian is very difficult and complicated. So we use another method to solve this problem.
We assume that the condensate wavefunction is in the form $\psi({\bf r})=\sqrt{N}\phi({\bf r}) $ and we can write the particle density as
\bea
n({\bf r})=|\psi({\bf r})|^{2}.
\label{the particle density}
\eea
The energy functional for the $N$-particle wavefunction $\phi$ is given by
\bea
E(\phi)=\frac{\langle\phi|\hat H|\phi\rangle}{\langle\phi| \phi \rangle}.
\label{the energy functional}
\eea
Also, the free energy $F$ is defined as follows
\bea
F=E-\mu N
\label{the free energy}
\eea
where $\mu$ is the chemical potential. Now we find a solution for the wave function by minimizing the free energy. Using the wave function, the free energy is evaluate  as follows
\bea
F(\psi)=\langle\psi|\hat H|\psi\rangle-\mu\langle\psi| \psi \rangle.
\label{free energy}
\eea
Thus, we calculate each of terms of Hamiltonian in Eq. (\ref{free energy}). For the first term of Eq.( \ref{HAMILTONIAN}); the Kinetic energy term, we find that
%\bea
%\langle {\psi}|  \sum_{i=1}^N \frac{{\bf p}^2}{2m} | {\psi}\rangle  &=& -N \frac{\hbar^2}{2m} \int \phi^*({\bf r}) D^2 \phi({\bf r}) d_{q}{\bf r},
%\label{HM}
%\eea
%Using Eq.(\ref{q-deform schrodinger equation}), we find that 
\bea
\langle {\psi}|  \sum_{i=1}^N \frac{{\bf p}^2}{2m} | {\psi}\rangle =
&-& N \frac{\hbar^2}{2m} \int \phi^*({\bf r}){\cal{D}}^{2(q)} \phi({\bf r}) d{\bf r}\nonumber\\
&=&-N \frac{\hbar^2}{2m} \int \phi^*({\bf r})[\frac{q}{(q-q^{-1})^{2}}(\frac{\psi(q^{-2}x)}{x^{2}}+\frac{\psi(q^{-2}y)}{y^{2}}+\frac{\psi(q^{-2}z)}{z^{2}})\nonumber\\
&-&\frac{q^{-1}}{(q-q^{-1})^2}(\frac{\psi(q^{2}x)}{x^2}+\frac{\psi(q^{2}y)}{y^2}+\frac{\psi(q^{2}z)}{z^2})
-\frac{1}{(q-q^{-1})}(\frac{\psi(x)}{x^2}+\frac{\psi(y)}{y^2}+\frac{\psi(z)}{z^2}]d{\bf r}.
\label{GG}
\eea
and the second term; the potential term, is written by
\bea
\langle{\psi}| \sum_{i=1}^N V_{ext}({\bf r}_i) |{\psi} \rangle= N \int\phi^*({\bf r}) V_{ext} \phi({\bf r}) d{\bf r}.
\label{HN}
\eea
Also, we find the third term of Eq. (\ref{HAMILTONIAN}) as follows
\bea
\langle {\psi}| \frac{1}{2} \sum_{i=1}^{N}\sum_{j\neq i}^N V\left(|{\bf r}_i-{\bf r}_j|\right) |{\psi} \rangle&=&\frac{1}{2}\sum_{i=1}^N \sum_{j \neq i}^N \int d{\bf r}_i \int \phi^*({\bf r}_i) \phi^*({\bf r}_j) V\left(|{\bf r}_i-{\bf r}_j|\right) \phi({\bf r}_i) \phi({\bf r}_j) d{\bf r}_j, \nonumber\\
  &=& \frac{N(N-1)}{2}\int d{\bf r} \int \phi^*({\bf r})\phi({\bf r}') V\left(|{\bf r}-{\bf r}'|\right) \phi({\bf r}) \phi({\bf r}') d{\bf r}'.
\label{HO}
\eea
Finally, the last term of Eq. (\ref{free energy}) is given by
\bea
 \mu\langle  {\psi}|{\psi}\rangle = \mu \left( \int \phi^*({\bf r}) \phi({\bf r}) d{\bf r} \right)^N.
\label{MU}
\eea
It has been shown that the thermodynamic curvature of deformed boson gas is singular at a critical value of fugacity as follows \cite{mirza2011thermodynamic}
\bea
 z_q=\left\{
         \begin{array}{cc}
           q^{2} & q<1 \\
           q^{-2} & q>1 \\
         \end{array}\right.\label{zq}.
\label{k}
\eea
In this paper, we restrict ourselves to the case that the deformation parameter belongs to $0\le q\le 1$. Thus, the system tends to the ordinary bosons  in the limit of $q\longrightarrow 1$ and also the small values of the deformation parameter equivalent to the more deformed cases. The chemical potential of particles of condensed phase with $q$- deformed boson statistics in $q<1$ will be as follows 
\bea
\mu=k_{B}T\ln q^{2}.
\label{fagucity}
\eea
%Now, we minimize the free energy using Eqs. (\ref{GG}),(\ref{HN}) and (\ref{HO})).
 We assume a small variation in the wavefunction $\phi({\bf r})$. The wavefunction is complex and have a real and an imaginary part. We consider $\phi$ and $\phi^*$ as the independent variables. Thus, we work out the functional derivatives of  Eqs. (\ref{GG}), (\ref{HN}) and (\ref{HO})) with respect to $\phi^*$ and obtain that
\bea
F(\psi)=&-&N \frac{\hbar^2}{2m} \int \phi^*({\bf r})\left[\frac{q}{(q-q^{-1})^{2}}\left(\frac{\phi(q^{-2}x)}{x^{2}}+\frac{\phi(q^{-2}y)}{y^{2}}+\frac{\phi(q^{-2}z)}{z^{2}}\right)
-\frac{q^{-1}}{(q-q^{-1})^{2}}\left(\frac{\phi(q^{2}x)}{x^{2}}+\frac{\phi(q^{2}y)}{y^{2}}+\frac{\phi(q^{2}z)}{z^{2}}\right)\right. \nonumber\\
&-&\left. \frac{1}{(q-q^{-1})}\left(\frac{\phi(x)}{x^{2}}+\frac{\phi(y)}{y^{2}}+\frac{\phi(z)}{z^{2}}\right)\right]d{\bf r}
+ N \int \phi^*({\bf r}) V_{ext} \phi({\bf r}) d{\bf r}\nonumber\\
&+& \frac{N(N-1)}{2}\int d{\bf r} \int \phi^*({\bf r})\phi({\bf r}') V\left(|{\bf r}-{\bf r}'|\right) \phi({\bf r}) \phi({\bf r}') d{\bf r}'
-k_{B}T\ln q^{2}\left( \int \phi^*({\bf r}) \phi({\bf r}) d{\bf r} \right)^N.
\label{f}
\eea
We minimize the free energy $\frac{\delta F}{\delta \phi^*}=0$, using the approximation $N-1\simeq N $ and $V(|{\bf r}-{{\bf r}'}|)=\frac{4\pi\hbar^2}{m} a\delta({\bf r}-{{\bf r}'})$ which is obtained as follows  
\bea
\frac{\delta F}{\delta \phi^* }=&-&N \frac{\hbar^2}{2m}\left[\left(\frac{q}{(q-q^{-1})^{2}}\frac{\phi(q^{-2}x)}{x^{2}}+\frac{\phi(q^{-2}y)}{y^{2}}+\frac{\phi(q^{-2}z)}{z^{2}} \right)
-\frac{q^{-1}}{(q-q^{-1})^{2}}\left(\frac{\phi(q^{2}x)}{x^{2}}+\frac{\phi(q^{2}y)}{y^{2}}+\frac{\phi(q^{2}z)}{z^{2}}\right)\right.\nonumber\\
&-&\left. \frac{1}{(q-q^{-1})}\left(\frac{\phi(x)}{x^{2}}+\frac{\phi(y)}{y^{2}}+\frac{\phi(z)}{z^{2}}\right)\right]
+N\left( V_{ext}({\bf r})  
+N\frac{4\pi\hbar^2}{m} a |\phi({\bf r})|^{2} 
-k_{B}T\ln q^{2} \right)\phi({\bf r})=0
\label{T}
\eea
According to $\psi({\bf r})=\sqrt{N}\phi({\bf r})$, Eq. (\ref{T}) becomes as follows 
\bea
&-& \frac{\hbar^2}{2m}\left[\left(\frac{q}{(q-q^{-1})^{2}}\frac{\psi(q^{-2}x)}{x^{2}}+\frac{\psi(q^{-2}y)}{y^{2}}+\frac{\psi(q^{-2}z)}{z^{2}}\right)
-\frac{q^{-1}}{(q-q^{-1})^{2}}\left(\frac{\psi(q^{2}x)}{x^{2}}+\frac{\psi(q^{2}y)}{y^{2}}+\frac{\psi(q^{2}z)}{z^{2}}\right) \right.\nonumber\\
&-&\left. \frac{1}{(q-q^{-1})}\left(\frac{\psi(x)}{x^{2}}+\frac{\psi(y)}{y^{2}}+\frac{\psi(z)}{z^{2}}\right)\right]
+\left( V_{ext}({\bf r})
+\frac{4\pi\hbar^2}{m} a |\psi({\bf r})|^{2} 
-k_{B}T\ln q^{2} \right)\psi({\bf r})=0
\label{GPE-Q-DEFORMED}
\eea
Eq. (\ref{GPE-Q-DEFORMED}) is the $q$-deformed the time-independent GPE. It is a kind of non-linear Schrodinger equation, that the total potential consists of the external potential $ V_{ext}({\bf r})$ and a non-linear term $\frac{4\pi\hbar^2}{m} a |\psi({\bf r})|^{2}\psi({\bf r})$ which describes the mean-field potential of the other atoms. 
In conclusion, the dynamics of identical particles system obeying deformed statistics can be described by the $q$-deformed GPE. This result is expected since in a system with ground state obeying deformed statistics the GPE should be different from a Bose-Einstein system.

%%%%%%%%%%%%%%%%%%%%%%%%%%%%%%%%%%%%%%%%%

%%%%%%%%%%%%%%%%%%%%%%%%%%%%%%%%%%%%%%%

%%%%%%%%%%%%%%%%%%%%%%%%%%%%%%%%%%%%%%%%

%%%%%%%%%%%%%%%%%%%%%%%%%%%%%%%%%%%%%%%%%%%%%%%%%%%%%%%%%%%%%%%%%%%

%%%%%%%%%%%%%%%%%%%%%%%%%%%%%%%%%%%%%%%%%
%%%%%%%%%%%%%%%%%%%%%%%%%%%%%%%%%%%%%%%%%%%%%%%%%%%%%

\section{Conclusion}\label{5}
GPE is a non linear equation that describes the Bose-Einstein condensate and has various utilization in various fields. For generalized statistics such as deformed statistics, fractional statistics and nonextensive statistics similar this equation is not yet known. Thus, calculation of this equation to describe the condensate of particles with generalized statistics could be interesting theoretically and might be useful in some aspects of physics. Of course, recently a generalized deformed GPE has been proposed to explain the properties of dark matter \cite{maleki2022description}. 

We considered the $q$-deformed formalism for the well-known Schrodinger equation. In the following, Using the free energy of a system of particles with deformed statistics in condensate phase, we calculate the $q$-deformed form of GPE which can also be useful for research on various topics in physics.
%%%%%%%%%%%%%%%%%%%%%%%%%%%%%%%%%%%%%%%%%%%%%%%%%%%%
 %%%%%%%%%%%%%%%%%%%%%%
%%%%%%%%%%%%%%%%%%%%%%%%%%%%%%%%%%%%%%%%
%%%%%%%%%%%%%%%%%%%%%%%%%%%%%%%%%%%%%%%%%%%%%%%%%%%%%%%%%
\bibliography{refs}
\end{document}